# Giant radiative thermal rectification using an intrinsic semiconductor film

Qizhang Li[1,2], Qun Chen[2] and Bai Song[1,3,4†]

[1]*Beijing Innovation Center for Engineering Science and Advanced Technology, Peking University, Beijing 100871, China*

[2]*Key Laboratory for Thermal Science and Power Engineering of Ministry of Education, Department of Engineering Mechanics, Tsinghua University, Beijing 100084, China*

[3]*Department of Energy and Resources Engineering, Peking University, Beijing 100871, China*

[4]*Department of Advanced Manufacturing and Robotics, Peking University, Beijing 100871, China*

[†]Corresponding author. Email: songbai@pku.edu.cn

**ABSTRACT:** Rectification of heat flow via a thermal diode is not only of fundamental interest, but can also enable a range of novel applications in thermal management and energy conversion. However, despite decades of extensive research, large rectification ratios of practical importance have yet to be demonstrated. Here, we theoretically achieve giant rectification ratios (3 to almost 5 orders of magnitude) by leveraging near-field radiative thermal transport between two parallel planes. Guided by a rational design approach centering on the electromagnetic local density of states (LDOS), we employ a thin film of an intrinsic semiconductor—such as silicon—as one terminal of our radiative thermal diodes, which provides the necessary nonlinearity and a substantial LDOS contrast as the temperature bias is flipped. For the other terminal, we explore two kinds of materials which either serve as a narrowband or a broadband filter, both capable of converting the large LDOS contrast into giant thermal rectification. We further consider representative multilayer configurations in order to block backside radiation and improve mechanical stability. All the diodes perform well over a wide range of film thicknesses, gap sizes, and temperatures. Our work offers an opportunity for realizing thermal diodes with unprecedented rectification ratios.

## INTRODUCTION

As a foundational building block of modern electronics, the electrical diode is a two-terminal device that enables the asymmetric transport of charges. In analogy, a thermal diode transports heat preferentially in one direction, and is of great interest to a variety of novel technologies in energy harvesting and thermal circuits.[1–3] Key to the performance of a thermal diode is its rectification ratio which is defined as $R = (Q_F - Q_R)/Q_R$, where $Q_F$ and $Q_R$ are the heat currents under a forward and a reverse temperature bias, respectively. Research on thermal rectification dates back at least to 1936, when Starr[4] experimentally observed an unusual asymmetric thermal conductance across a copper-cuprous oxide interface, with rectification ratios around 1. A notable theoretical advancement was made in 2002, when Terraneo *et al.*[5] showed that tailored 1D nonlinear chains could potentially rectify heat current by roughly a factor of 2. Over the past two decades, thermal rectification mediated by phonons,[5–10] electrons,[11–13] and photons[14–20] have been extensively studied. However, to date most experimental efforts have only realized rectification ratios around 1,[7–10,17] and much larger values of practical importance remain to be demonstrated.

Recently, near-field radiative thermal diodes exploiting the tunneling of various evanescent electromagnetic waves have attracted broad attention, because they have the potential to achieve ultrahigh rectification over a wide temperature range.[21–27] Theoretically, a plethora of designs featuring polar dielectrics,[14,28,29] phase transition materials,[25–27,30–32] and semiconductors[22,33–36] have been proposed, with the rectification ratios gradually increasing from 0.4 to over 324.[14,27] Experimental demonstration, however, lags far behind.[17] The key challenge lies, first of all, in the design of radiative diodes[25] that combine large rectification with the ease of fabrication and characterization.[37–42]

In this work, we theoretically demonstrate giant rectification ratios in a suite of radiative thermal diodes in the parallel-plane geometry, which is technologically the most promising configuration. At the heart of our diodes is a proper intrinsic semiconductor which provides the necessary nonlinearity[43] due to its strongly temperature-dependent free-carrier concentration and infrared permittivity.[44,45] To rationally design for maximum rectification, we employ an approach centering on the electromagnetic local density of states (LDOS).[25,27] The intrinsic semiconductor offers a large LDOS contrast especially at low frequency as the temperature bias is flipped. To translate this LDOS contrast into rectification, we explore two kinds of bandpass filters to pair with the semiconductor: one limits the thermal transport to a desired narrow spectral range with its sharp LDOS peak, while the other features large LDOS over a broad spectral range.

As an example, we consider intrinsic silicon which is arguably the most fabrication-friendly semiconductor with well-characterized optical properties. We achieve rectification ratios on the order of $10^3$ and $10^5$ at around 100-nm gaps, by pairing intrinsic-silicon thin films with films of potassium bromide and doped silicon, which serve as the narrowband and broadband filter, respectively. To block backside radiation and improve mechanical stability, we further propose a multilayer configuration with highly-reflective metal substrates such as silver, and achieve a rectification ratio of 709 (1777) at a gap of 10 nm (5 nm). For all the diodes, ultrahigh rectification ratios ($R > 100$) can be obtained over a wide range of film thicknesses, gap sizes, and temperatures.

**RESULTS AND DISCUSSION**

In order to explore the rectification potential of semiconductors, we employ an approach based on the contrast of near-field LDOS arising from their temperature-dependent permittivity.[25,27] For a semiconductor, the permittivity essentially follows the Drude model as[44,46]

$$\varepsilon(\omega) = \varepsilon_{bl} - \frac{N_e e^2 / \varepsilon_0 m_e^*}{\omega^2 + i\omega/\tau_e} - \frac{N_h e^2 / \varepsilon_0 m_h^*}{\omega^2 + i\omega/\tau_h}, \tag{1}$$

where $\varepsilon_{bl}$ represents the contributions from interband transitions and lattice absorption; $N_e$, $m_e^*$, and $\tau_e$ are the concentration, effective mass, and relaxation time, respectively, of the free electrons in the conduction band; $N_h$, $m_h^*$, and $\tau_h$ are the corresponding quantities for the holes in the valence band; $\varepsilon_0$ is the permittivity of free space and $\omega$ is the angular frequency. As temperature rises, more free carriers are thermally generated which leads to a large increase in the imaginary part of the permittivity ($\varepsilon''$) especially in the infrared, while the real part ($\varepsilon'$) is usually much less affected.[44,46] Below we focus on intrinsic semiconductors which are more sensitive to temperature variations than doped ones.

To go from the temperature-dependent permittivity to the characteristics of the LDOS, we utilize the generalized emissivity of the evanescent modes, Im($r$),[47,48] which is proportional to the wavevector-resolved local density of states (WLDOS).[49–52] We consider the high-$\kappa$ modes with large parallel wavevectors since they usually dominate near-field radiation.[48,53,54] For a bulk material, the emissivity of the $s$-polarized high-$\kappa$ modes is close to zero, while that of the $p$-polarized modes is

$$\text{Im}(r) \approx \frac{2\varepsilon''}{(\varepsilon'+1)^2 + \varepsilon''^2} \approx \frac{2\varepsilon''}{(\varepsilon'+1)^2}, \tag{2}$$

since $\varepsilon''$ of an intrinsic semiconductor in the infrared is often negligibly small compared to $|\varepsilon'|$. According to Eq. (2), the distinct temperature dependence of $\varepsilon''$ and $\varepsilon'$ for an intrinsic semiconductor would then lead to a large emissivity variation for the $p$-polarized high-$\kappa$ modes as temperature varies, which potentially results in a large LDOS contrast and forms the basis for ultrahigh thermal rectification.

As an example, we quantitatively analyze the permittivity and LDOS characteristics of intrinsic silicon (*i*-Si) (Fig. 1). Here, the temperature-dependent permittivity of *i*-Si is mainly obtained from Ref. 46, albeit with a modification of $\varepsilon_{bl}$ according to various experimental measurements,[55–59] which now properly incorporates the lattice absorption over a very broad frequency range from $10^{12}$ rad s$^{-1}$ to about $2\times10^{15}$ rad s$^{-1}$. Note that neglection of the broadband lattice absorption could exaggerate the rectification ratios of *i*-Si-based diodes by orders of magnitude [see Supporting Information (SI) for details].

For *i*-Si at 300 K, $\varepsilon''$ is dominated by lattice absorption, which is overwhelmed at higher temperatures by contribution from the free carriers (Fig. 1a). As the temperature increases from 300 K to 1000 K, the $\varepsilon''$ of *i*-Si increases by three to four orders of magnitude, while the increase of $\varepsilon'$ is less than 12%. In Fig. 1b, we plot the corresponding LDOS at 100 nm above bulk *i*-Si, which shows distinct temperature-dependent characteristics in different spectral ranges. At high frequency, the LDOS is mainly contributed by the propagating modes within the medium ($\kappa <$ Re$(\sqrt{\varepsilon})k_0$), which remains almost constant regardless of the varying temperatures due to the temperature-insensitive $\varepsilon'$. In comparison, the LDOS at low frequency is dominated by the *p*-polarized high-$\kappa$ modes, and is highly sensitive to temperature variations due to the dramatic changes in $\varepsilon''$. This is qualitatively consistent with Eq. (2) and is quantitatively confirmed by calculations in Fig. 1c.

Although a large LDOS contrast is readily established at low frequency, the high-frequency propagating modes in bulk *i*-Si remain nearly unaffected by the varying temperature (Fig. 2a-c), and thus can severely limit the rectification ratio. These modes can be largely suppressed by employing a thin film of *i*-Si instead of a bulk due to the much smaller source volume (Fig. 2d-f).

[25,60] Meanwhile, the low-frequency high-$\kappa$ modes localized at the *i*-Si – vacuum interface are only slightly affected. The dramatic contrast of low-frequency LDOS exhibited by the *i*-Si film (Fig. 2g), if properly utilized, can lead to extraordinarily large thermal rectification.

To convert the LDOS contrast into thermal rectification, we first pair an *i*-Si film with a narrow bandpass filter—a polar dielectric which features a sharp LDOS peak due to its surface phonon polaritons (SPhPs) (Fig. 3).[25] Many polar dielectrics can potentially be used, such as cubic boron nitride (cBN),[61] silicon carbide (SiC),[53] barium fluoride ($BaF_2$),[55] potassium chloride (KCl),[55] and potassium bromide (KBr).[55] Among these materials, we focus on KBr because its LDOS peak is located around a frequency ($\omega_{KBr} = 2.7 \times 10^{13}$ rad s$^{-1}$) where the *i*-Si film features the highest LDOS contrast (Fig. 3a).

With the terminal temperatures of the thermal diodes set at $T_{low}$ = 300 K and $T_{high}$ = 1000 K, we calculate the forward and reverse heat fluxes for the *i*-Si – KBr pair in three configurations at a 100-nm gap (Fig. 3b and Fig. S3), based on the theoretical framework of fluctuational electrodynamics.[48,62] We first consider the configuration of bulk *i*-Si paired with bulk KBr, which shows a negligibly small rectification ratio ($R$ = 0.03) because both the forward and reverse fluxes are dominated by high-frequency contributions (>94% from $10^{14} - 2 \times 10^{15}$ rad s$^{-1}$) which are insensitive to the direction of the temperature bias. As soon as we replace the bulk *i*-Si with a 10-nm-thick film, the rectification ratio increases sharply to 1673 (>50000× enhancement), since the high-frequency contributions are now greatly suppressed (<4%). The rectification ratio further increases to 2517 when the bulk KBr is also replaced with a 10-nm-thick film, which features coupled SPhPs and a LDOS peak redshifted[25,63] to the spectral range where the *i*-Si film shows a larger LDOS contrast (Fig. 3a,b).

In Fig. 3c, we show a map of the rectification ratio with respect to the thicknesses of the $i$-Si and KBr films, at a gap size of 100 nm. Rectification ratios greater than 1600 can be achieved over a wide thickness range (1 nm – 1 μm) for both films. With a 1000-fold increase in the $i$-Si film thickness, the rectification only decreases by less than 13%, which is expected from the slight variance of the LDOS contrast (Fig. 2g and Fig. 3c). This characteristic of the $i$-Si film is quite different from that of a thin film of metal-to-insulator transition materials, which features a highly thickness-dependent LDOS contrast due to the coupling of surface modes within the film.[25] For $i$-Si, the LDOS contrast at low frequency results from the high-$\kappa$ modes at high temperatures (e.g., 1000 K) which are largely insensitive to the film thickness (Fig. 2). As for the KBr film, the decreasing thickness leads to a larger redshift of the LDOS peak and thus a monotonous increase of rectification. We note that similarly large rectification ratios can also be achieved by using KCl or $BaF_2$ as the narrowband filter (Fig. S4), both of which feature LDOS peaks at relatively low frequencies where $i$-Si shows a substantial LDOS contrast.

Despite the effectiveness of using polar dielectrics as narrowband filters to achieve ultrahigh rectification, this approach is fundamentally limited by the availability of materials with the desired SPhPs. In order to take full advantage of the even larger LDOS contrast of the $i$-Si film over the wide spectral range below $\omega_{\mathrm{KBr}}$ (Fig. 3a), we now explore the potential of a broadband filter which ideally is characterized by a large LDOS over a wide range of low frequencies. As an example, we propose that doped silicon ($d$-Si) can be a good candidate (Fig. 4 and Fig. S5). We first consider $n$-type $d$-Si with a doping level ($N_d$) of $10^{18}$ cm$^{-3}$, which is close to the free-carrier concentration of $i$-Si at 1000 K and likely to yield a large heat flux in the forward scenario. Similar to $i$-Si at high temperatures, bulk $d$-Si features large LDOS at both low and high frequencies (Fig.

4a), as dictated by the high-$\kappa$ modes and the propagating modes in *d*-Si, respectively. These high-$\kappa$ modes survive in a thin film of *d*-Si, while the propagating modes are consistently suppressed with decreasing film thickness (Fig. 4a).

By pairing a 10-nm-thick *i*-Si film with a *d*-Si film of the same thickness, we obtain a forward heat flux that is much larger than the reverse flux almost over the entire spectral range of interest (Fig. 4b). This leads to a giant rectification ratio of 12654 at a 100-nm gap, more than $2\times10^6$ times larger than that of the bulk-bulk case ($R = 0.005$). Compared to the case of *i*-Si film paired with bulk *d*-Si ($R = 4665$), the rectification increase in the film-film case mainly originates from a 3.5-fold enhancement of the forward heat flux, with dominant contributions from frequencies below $2\times10^{13}$ rad s$^{-1}$. Here, the low-frequency heat flux is enhanced by the coupling of SPPs within the *d*-Si film at 300 K, which leads to enhanced LDOS (Fig. 4a)[64,65] and also accounts for the increasing rectification with decreasing *d*-Si film thickness (Fig. 4c). With the diode combining thin films of *i*-Si and *d*-Si, rectification ratios of over $10^4$ can be obtained over wide thickness ranges (Fig. 4c). Compared to the diode using KBr, the rectification with *d*-Si decreases much more rapidly when the film thickness of *i*-Si is larger than ~100 nm. This is because *i*-Si and *d*-Si have similarly large $\varepsilon'$, so that the contribution from the modes propagating in *i*-Si would easily increase the reverse heat flux for a relatively thick *i*-Si film (Fig. 4c). By optimizing the thicknesses of both films and the doping level of *d*-Si, we achieve a record-high rectification ratio of 52409 at a gap of 100 nm, and 85888 at a 200-nm gap (Table S1), which highlights the great potential of parallel-plane near-field radiative thermal diodes.

To further improve the experimental feasibility of *i*-Si-based radiative thermal rectification, we now propose a few multilayer configurations with highly reflective substrates capable of blocking the backside radiation. This would allow the diodes to be supported on any medium without affecting its rectification performance. Here, we employ a silver (Ag) substrate, and begin with an analysis of the LDOS characteristics for the *i*-Si side. The permittivity of Ag is from Ref. 66 with a temperature-dependent scattering rate.[34,67] We first consider the configuration with a 10-nm-thick *i*-Si film directly sitting on a bulk Ag substrate. Unfortunately, the LDOS contrast at $z$ = 100 nm above the *i*-Si/Ag bilayer is orders-of-magnitude smaller than that of a suspended *i*-Si film of the same thickness (Fig. 5a), due to the temperature-insensitive *s*-polarized modes introduced by the Ag substrate (Fig. S6). In order to increase the LDOS contrast for *i*-Si/Ag, we make use of two geometric factors. First, the *i*-Si film should be sufficiently thick to distance the Ag substrate from the vacuum interface, but not too thick to introduce a noticeable contribution from the propagating modes inside the *i*-Si. Further, the distance $z$ needs to be small enough to ensure that the high-$\kappa$ modes dominate the LDOS, since the density of these modes varies more rapidly with $z$ than the *s*-polarized modes.[54,68] By using a 100-nm-thick *i*-Si film, a much larger LDOS contrast is obtained at $z$ =10 nm above the *i*-Si/Ag bilayer over a broad spectral range (Fig. 5a).

To see how the Ag substrate affects the rectification performance, we first analyze the diode with a narrowband filter (Fig. 5b and Fig. S7a). For the pair of *i*-Si/Ag and KBr/Ag at a gap size of 10 nm, we achieve an optimized rectification of 397 when the thickness of the KBr film is 82 μm, much larger than that of the *i*-Si film (65 nm). This is because KBr has a small $\varepsilon'$ so that a thick film can keep the Ag substrate far away from the vacuum gap without introducing much reverse heat flux. However, for the diode consisting of *i*-Si/Ag and *d*-Si/Ag (Fig. 5c and Fig. S7b), the optimized rectification ratio is only 80, much smaller than that achieved with suspended *i*-Si and *d*-Si films ($R_{opt}$ = 14266 at a 10-nm gap). This is because, in order to separate the Ag layer from the gap, the *d*-Si layer itself has to be very thick, which introduces much heat flux due to the propagating modes in *d*-Si (Fig. S7b). To suppress this undesired flux, we propose an improved configuration by inserting a KBr layer between the *d*-Si film and the Ag substrate (Fig. 5d).[25] When paired with *i*-Si/Ag, this trilayer structure yields an optimized rectification ratio of 709 at a gap of 10 nm, which increases to 1777 at a 5-nm gap (Table S1). In this diode, the transparent KBr interlayer also allows coupled SPPs within the *d*-Si film at 300 K, which accounts for the increase of rectification with decreasing *d*-Si thickness (Fig. 5e) as discussed above. We note that, since the *d*-Si film has a large LDOS over a wide spectral range, polar dielectrics supporting SPhPs should not be used as an interlayer on the *i*-Si side in order to avoid a large reverse heat flux.

With the basic configurations and characteristics of our radiative thermal diodes presented, we now proceed to systematically analyzing the effects of the gap size and temperature. Specifically, we consider three representative diodes including *i*-Si film paired with KBr film, *i*-Si film paired with *d*-Si film, and the trilayer configuration with Ag substrates (Fig. 6). For the two diodes with suspended films, the rectification ratios first increase with increasing gap size. This is because a larger gap leads to a larger penetration depth of the resonant surface modes in the KBr or the *d*-Si film,[40,69] and consequently a redshift of the corresponding LDOS peak. At gaps larger than 100 nm, the rapid decay of the high-$\kappa$ modes with increasing gap size explains the decreasing rectification ratios. In comparison, the rectification ratio of the multilayer diode monotonically increases towards smaller gaps, since the high-$\kappa$ *p*-polarized modes become increasingly dominant over the *s*-polarized modes induced by the Ag substrates. For the suspended-film diodes, rectification ratios larger than $10^3$ can be achieved at gap sizes up to at least 1 μm. When optimized at each gap size, all the diodes feature ultrahigh rectification ratios (> 100) over a broad gap-size range even for a much smaller $T_{high}$ of 700 K.

In Fig. 6b, we show the heat fluxes under different temperature biases ($\Delta T$) for the suspended-film and multilayer diodes, all exhibiting a remarkable nonlinearity. As the temperature bias increases, these diodes yield higher rectification ratios since more free carriers are thermally excited in the *i*-Si at $T_{high}$ which lead to a larger LDOS contrast. For all the diodes, over 100-fold rectification can be obtained over a wide temperature range ($|\Delta T| > 400$ K). Given the recent success of measuring near-field radiative heat transfer under a temperature difference of over 900 K,[70] we are hopeful that the giant thermal rectification predicted in this work could be experimentally realized in the near future.

**CONCLUSIONS**

In summary, we have systematically explored the rectification potential of intrinsic semiconductors. With intrinsic silicon as an example, we theoretically achieve three to almost five orders of magnitude radiative thermal rectification in the parallel-plane geometry over a wide temperature range, which represents 100 to 1000-fold improvement over state-of-the-art designs (Fig. S9). Key to achieving the giant thermal rectification is the ultrahigh LDOS contrast provided by a thin film of intrinsic silicon, which is combined with either a narrowband or a broadband filter. Representative multilayer configurations incorporating highly-reflective substrates are also analyzed to block backside radiation and improve mechanical stability. We expect these promising results to roughly apply to other intrinsic semiconductors such as germanium, which share similar temperature-dependent permittivities.[44,55,71] Our work reveals new possibilities for the long-awaited experimental realization of a thermal diode with giant rectification.

**METHODS**

The LDOS at a distance $z$ above a planar medium can be calculated by[53]

$$\rho(\omega, z) = \int_0^\infty \left[ \rho_{w,s}(\kappa, \omega, z) + \rho_{w,p}(\kappa, \omega, z) \right] d\kappa, \quad (3)$$

where $\rho_{w,s}(\omega, \kappa, z)$ and $\rho_{w,p}(\omega, \kappa, z)$ are the WLDOS for the *s*- and *p*-polarized modes, respectively, and can be expressed in terms of the Fresnel reflection coefficients ($r$) as follows:

$$\rho_{w,\alpha=s,p}(\omega, \kappa, z) = \begin{cases} \dfrac{\omega^2}{4\pi^2 c^3} \dfrac{\kappa}{k_0 |\gamma_0|} \left(1 - |r_\alpha|^2\right), & \kappa \leq k_0, \\[2mm] \dfrac{\omega^2}{2\pi^2 c^3} \dfrac{\kappa^3}{k_0^3 |\gamma_0|} \mathrm{Im}(r_\alpha) e^{-2\mathrm{Im}(\gamma_0)z}, & \kappa > k_0, \end{cases} \quad (4)$$

where $\gamma_0$ is the normal wavevector in vacuum and $k_0 = \omega/c$.

Based on the theoretical framework of fluctuational electrodynamics, the radiative heat flux between two planes across a vacuum gap $d$ is given by[62]

$$q(T_1, T_2, d) = \int_0^\infty \frac{d\omega}{4\pi^2} \left[ \Theta(\omega, T_1) - \Theta(\omega, T_2) \right] \int_0^\infty d\kappa \kappa \left[ \tau_s^{12}(\omega, \kappa) + \tau_p^{12}(\omega, \kappa) \right], \quad (5)$$

where $\Theta(\omega, T) = \dfrac{\hbar \omega}{\exp(\hbar\omega/k_B T) - 1}$ is the mean energy of a harmonic oscillator at temperature $T$; $\tau_s^{12}(\omega, \kappa)$ and $\tau_p^{12}(\omega, \kappa)$ are respectively the transmission probabilities for the *s*- and *p*-polarized modes, which can be obtained as

$$\tau_{\alpha=s,p}^{12}(\omega, \kappa) = \begin{cases} \dfrac{\left(1 - |r_\alpha^1|^2\right)\left(1 - |r_\alpha^2|^2\right)}{\left|1 - r_\alpha^1 r_\alpha^2 \exp(2i\gamma_0 d)\right|^2}, & \kappa \leq k_0, \\[2mm] \dfrac{4\,\mathrm{Im}(r_\alpha^1)\,\mathrm{Im}(r_\alpha^2) e^{-2\mathrm{Im}(\gamma_0)d}}{\left|1 - r_\alpha^1 r_\alpha^2 \exp(2i\gamma_0 d)\right|^2}, & \kappa > k_0. \end{cases} \quad (6)$$

For suspended thin films, $1 - |r_\alpha|^2$ in Eq. (4) should be replaced by $1 - |r_\alpha|^2 - |t_\alpha|^2$, where $t_\alpha$ is the transmission coefficient. In the optimization process to find the maximum rectification, the lower and upper limits of film thickness are respectively set to 1 nm and 1 cm, and that of $N_d$ are $10^{15}$ cm$^{-3}$ and $10^{20}$ cm$^{-3}$, respectively.

## ACKNOWLEDGEMENT

This work is supported by the National Natural Science Foundation of China (Grants No. 52076002 and No. 51836004), the Beijing Innovation Center for Engineering Science and Advanced Technology (BIC-ESAT), the XPLORER PRIZE from the Tencent Foundation, and the High-performance Computing Platform of Peking University.

## SUPPORTING INFORMATION

Permittivity of *i*-Si considering broadband lattice absorption, transmission probabilities for various diodes pairing *i*-Si with KBr, rectification map for the diodes paring *i*-Si film with KCl film or BaF$_2$ film, transmission probabilities for the diodes pairing *i*-Si with *d*-Si, WLDOS for the *i*-Si/Ag bilayer at 300 K and 1000 K, rectification map for the diodes pairing *i*-Si/Ag with KBr/Ag or *d*-Si/Ag, permittivity of *d*-Si, KBr, KCl, BaF$_2$, and Ag, and comparison with state-of-the-art designs.

## REFERENCES

(1) Li, N.; Ren, J.; Wang, L.; Zhang, G.; Hänggi, P.; Li, B. Colloquium: Phononics: Manipulating Heat Flow with Electronic Analogs and Beyond. *Rev. Mod. Phys.* **2012**, *84* (3), 1045–1066.

(2) Wehmeyer, G.; Yabuki, T.; Monachon, C.; Wu, J.; Dames, C. Thermal Diodes, Regulators, and Switches: Physical Mechanisms and Potential Applications. *Appl. Phys. Rev.* **2017**, *4* (4), 041304.

(3) Li, Y.; Li, W.; Han, T.; Zheng, X.; Li, J.; Li, B.; Fan, S.; Qiu, C.-W. Transforming Heat Transfer with Thermal Metamaterials and Devices. *Nat. Rev. Mater.* **2021**, *6* (6), 488–507.

(4) Starr, C. The Copper Oxide Rectifier. *J. Appl. Phys.* **1936**, *7* (1), 15–19.

(5) Terraneo, M.; Peyrard, M.; Casati, G. Controlling the Energy Flow in Nonlinear Lattices: A Model for a Thermal Rectifier. *Phys. Rev. Lett.* **2002**, *88* (9), 094302.

(6) Li, B.; Wang, L.; Casati, G. Thermal Diode: Rectification of Heat Flux. *Phys. Rev. Lett.*

**2004**, *93* (18), 184301.

(7) Chang, C. W.; Okawa, D.; Majumdar, A.; Zettl, A. Solid-State Thermal Rectifier. *Science* **2006**, *314*, 1121.

(8) Wang, H.; Hu, S.; Takahashi, K.; Zhang, X.; Takamatsu, H.; Chen, J. Experimental Study of Thermal Rectification in Suspended Monolayer Graphene. *Nat. Commun.* **2017**, *8* (1), 15843.

(9) Kasprzak, M.; Sledzinska, M.; Zaleski, K.; Iatsunskyi, I.; Alzina, F.; Volz, S.; Sotomayor Torres, C. M.; Graczykowski, B. High-Temperature Silicon Thermal Diode and Switch. *Nano Energy* **2020**, *78*, 105261.

(10) Shrestha, R.; Luan, Y.; Luo, X.; Shin, S.; Zhang, T.; Smith, P.; Gong, W.; Bockstaller, M.; Luo, T.; Chen, R.; Hippalgaonkar, K.; Shen, S. Dual-Mode Solid-State Thermal Rectification. *Nat. Commun.* **2020**, *11* (1), 4346.

(11) Scheibner, R.; König, M.; Reuter, D.; Wieck, A. D.; Gould, C.; Buhmann, H.; Molenkamp, L. W. Quantum Dot as Thermal Rectifier. *New J. Phys.* **2008**, *10* (8), 083016.

(12) Ruokola, T.; Ojanen, T. Single-Electron Heat Diode: Asymmetric Heat Transport between Electronic Reservoirs through Coulomb Islands. *Phys. Rev. B* **2011**, *83* (24), 241404.

(13) Martínez-Pérez, M. J.; Fornieri, A.; Giazotto, F. Rectification of Electronic Heat Current by a Hybrid Thermal Diode. *Nat. Nanotechnol.* **2015**, *10* (4), 303–307.

(14) Otey, C. R.; Lau, W. T.; Fan, S. Thermal Rectification through Vacuum. *Phys. Rev. Lett.* **2010**, *104* (15), 154301.

(15) Ben-Abdallah, P.; Biehs, S. A. Phase-Change Radiative Thermal Diode. *Appl. Phys. Lett.* **2013**, *103* (19), 191907.

(16) Nefzaoui, E.; Drevillon, J.; Ezzahri, Y.; Joulain, K. Simple Far-Field Radiative Thermal Rectifier Using Fabry-Perot Cavities Based Infrared Selective Emitters. *Appl. Opt.* **2014**, *53* (16), 3479–3485.

(17) Fiorino, A.; Thompson, D.; Zhu, L.; Mittapally, R.; Biehs, S.-A.; Bezencenet, O.; El-Bondry, N.; Bansropun, S.; Ben-Abdallah, P.; Meyhofer, E.; Reddy, P. A Thermal Diode Based on Nanoscale Thermal Radiation. *ACS Nano* **2018**, *12* (6), 5774–5779.

(18) Marchegiani, G.; Braggio, A.; Giazotto, F. Highly Efficient Phase-Tunable Photonic Thermal Diode. *Appl. Phys. Lett.* **2021**, *118* (2), 022602.

(19) Sarkar, S.; Nefzaoui, E.; Basset, P.; Bourouina, T. Far-Field Radiative Thermal Rectification with Bulk Materials. *J. Quant. Spectrosc. Radiat. Transf.* **2021**, *266*, 107573.

(20) Latella, I.; Ben-Abdallah, P.; Nikbakht, M. Radiative Thermal Rectification in Many-Body Systems. *Phys. Rev. B* **2021**, *104* (4), 045410.

(21) Zhu, L.; Otey, C. R.; Fan, S. Ultrahigh-Contrast and Large-Bandwidth Thermal Rectification in near-Field Electromagnetic Thermal Transfer between Nanoparticles. *Phys. Rev. B* **2013**, *88* (18), 184301.

(22) Wen, S.; Liu, X.; Cheng, S.; Wang, Z.; Zhang, S.; Dang, C. Ultrahigh Thermal Rectification Based on Near-Field Thermal Radiation between Dissimilar Nanoparticles. *J. Quant. Spectrosc. Radiat. Transf.* **2019**, *234*, 1–9.

(23) Ott, A.; Biehs, S.-A. Thermal Rectification and Spin-Spin Coupling of Nonreciprocal Localized and Surface Modes. *Phys. Rev. B* **2020**, *101* (15), 155428.

(24) Zhang, Y.; Zhou, C.-L.; Yi, H.-L.; Tan, H.-P. Radiative Thermal Diode Mediated by Nonreciprocal Graphene Plasmon Waveguides. *Phys. Rev. Appl.* **2020**, *13* (3), 034021.

(25) Li, Q.; He, H.; Chen, Q.; Song, B. Thin-Film Radiative Thermal Diode with Large Rectification. *Phys. Rev. Appl.* **2021**, *16* (1), 014069.

(26) Moncada-Villa, E.; Cuevas, J. C. Normal-Metal--Superconductor Near-Field Thermal Diodes and Transistors. *Phys. Rev. Appl.* **2021**, *15* (2), 024036.

(27) Li, Q.; He, H.; Chen, Q.; Song, B. Radiative Thermal Diode via Hyperbolic Metamaterials. *Phys. Rev. Appl.* **2021**, In press.

(28) Joulain, K.; Ezzahri, Y.; Drevillon, J.; Rousseau, B.; De Sousa Meneses, D. Radiative Thermal Rectification between SiC and $SiO_2$. *Opt. Express* **2015**, *23* (24), A1388–A1397.

(29) Tang, L.; Francoeur, M. Photonic Thermal Diode Enabled by Surface Polariton Coupling in Nanostructures. *Opt. Express* **2017**, *25* (24), A1043–A1052.

(30) Yang, Y.; Basu, S.; Wang, L. Radiation-Based near-Field Thermal Rectification with Phase Transition Materials. *Appl. Phys. Lett.* **2013**, *103* (16), 163101.

(31) Huang, J.; Li, Q.; Zheng, Z.; Xuan, Y. Thermal Rectification Based on Thermochromic Materials. *Int. J. Heat Mass Transf.* **2013**, *67*, 575–580.

(32) Ghanekar, A.; Ji, J.; Zheng, Y. High-Rectification near-Field Thermal Diode Using Phase Change Periodic Nanostructure. *Appl. Phys. Lett.* **2016**, *109* (12), 123106.

(33) Basu, S.; Francoeur, M. Near-Field Radiative Transfer Based Thermal Rectification Using Doped Silicon. *Appl. Phys. Lett.* **2011**, *98* (11), 113106.

(34) Wang, L. P.; Zhang, Z. M. Thermal Rectification Enabled by Near-Field Radiative Heat Transfer Between Intrinsic Silicon and a Dissimilar Material. *Nanoscale Microscale Thermophys. Eng.* **2013**, *17* (4), 337–348.

(35) Shen, J.; Liu, X.; He, H.; Wu, W.; Liu, B. High-Performance Noncontact Thermal Diode via Asymmetric Nanostructures. *J. Quant. Spectrosc. Radiat. Transf.* **2018**, *211*, 1–8.

(36) Zhou, C. L.; Qu, L.; Zhang, Y.; Yi, H. L. Fabry-Perot Cavity Amplification of near-Field Thermal Rectification. *J. Quant. Spectrosc. Radiat. Transf.* **2020**, *251*, 107023.

(37) Shen, S.; Narayanaswamy, A.; Chen, G. Surface Phonon Polaritons Mediated Energy Transfer between Nanoscale Gaps. *Nano Lett.* **2009**, *9* (8), 2909–2913.

(38) Rousseau, E.; Siria, A.; Jourdan, G.; Volz, S.; Comin, F.; Chevrier, J.; Greffet, J.-J. Radiative Heat Transfer at the Nanoscale. *Nat. Photonics* **2009**, *3* (9), 514–517.

(39) Kim, K.; Song, B.; Fernández-Hurtado, V.; Lee, W.; Jeong, W.; Cui, L.; Thompson, D.; Feist, J.; Reid, M. T. H.; García-Vidal, F. J.; Cuevas, J. C.; Meyhofer, E.; Reddy, P. Radiative Heat Transfer in the Extreme near Field. *Nature* **2015**, *528* (7582), 387–391.

(40) Song, B.; Ganjeh, Y.; Sadat, S.; Thompson, D.; Fiorino, A.; Fernández-Hurtado, V.; Feist, J.; Garcia-Vidal, F. J.; Cuevas, J. C.; Reddy, P.; Meyhofer, E. Enhancement of Near-Field Radiative Heat Transfer Using Polar Dielectric Thin Films. *Nat. Nanotechnol.* **2015**, *10* (3), 253–258.

(41) Song, B.; Thompson, D.; Fiorino, A.; Ganjeh, Y.; Reddy, P.; Meyhofer, E. Radiative Heat Conductances between Dielectric and Metallic Parallel Plates with Nanoscale Gaps. *Nat. Nanotechnol.* **2016**, *11* (6), 509–514.

(42) Fiorino, A.; Thompson, D.; Zhu, L.; Song, B.; Reddy, P.; Meyhofer, E. Giant Enhancement in Radiative Heat Transfer in Sub-30 nm Gaps of Plane Parallel Surfaces. *Nano Lett.* **2018**, *18* (6), 3711–3715.

(43) Maznev, A. A.; Every, A. G.; Wright, O. B. Reciprocity in Reflection and Transmission: What Is a 'Phonon Diode'? *Wave Motion* **2013**, *50* (4), 776–784.

(44) Timans, P. J. *Advances in Rapid Thermal and Integrated Processing*; Kluwer Academic Publishers, Netherlands, 1996.

(45) A.Neamen, D. *Semiconductor Physics and Devices*; McGraw-Hill, New York, 2003.

(46) Fu, C. J.; Zhang, Z. M. Nanoscale Radiation Heat Transfer for Silicon at Different Doping Levels. *Int. J. Heat Mass Transf.* **2006**, *49*, 1703–1718.

(47) Mulet, J.-P.; Joulain, K.; Carminati, R.; Greffet, J.-J. Enhanced Radiative Heat Transfer at Nanometric Distances. *Microscale Therm. Eng.* **2002**, *6* (3), 209-222.

(48) Biehs, S.-A.; Rousseau, E.; Greffet, J.-J. Mesoscopic Description of Radiative Heat Transfer at the Nanoscale. *Phys. Rev. Lett.* **2010**, *105* (23), 234301.

(49) Pendry, J. B. Radiative Exchange of Heat between Nanostructures. *J. Phys. Condens. Matter* **1999**, *11* (35), 6621–6633.

(50) Guo, Y.; Cortes, C. L.; Molesky, S.; Jacob, Z. Broadband Super-Planckian Thermal Emission from Hyperbolic Metamaterials. *Appl. Phys. Lett.* **2012**, *101* (13), 131106.

(51) Guo, Y.; Jacob, Z. Thermal Hyperbolic Metamaterials. *Opt. Express* **2013**, *21* (12), 15014–15019.

(52) Liu, B.; Shen, S. Broadband Near-Field Radiative Thermal Emitter/Absorber Based on Hyperbolic Metamaterials: Direct Numerical Simulation by the Wiener Chaos Expansion Method. *Phys. Rev. B* **2013**, *87* (11), 115403.

(53) Joulain, K.; Mulet, J.-P.; Marquier, F.; Carminati, R.; Greffet, J.-J. Surface Electromagnetic Waves Thermally Excited: Radiative Heat Transfer, Coherence Properties and Casimir Forces Revisited in the near Field. *Surf. Sci. Rep.* **2005**, *57* (3), 59–112.

(54) Biehs, S.-A.; Tschikin, M.; Ben-Abdallah, P. Hyperbolic Metamaterials as an Analog of a Blackbody in the Near Field. *Phys. Rev. Lett.* **2012**, *109* (10), 104301.

(55) Palik, E. D. *Handbook of Optical Constants of Solids*; Elsevier, New York, 1985.

(56) Grischkowsky, D.; Keiding, S.; van Exter, M.; Fattinger, C. Far-Infrared Time-Domain Spectroscopy with Terahertz Beams of Dielectrics and Semiconductors. *J. Opt. Soc. Am. B* **1990**, *7* (10), 2006–2015.

(57) Green, M. A.; Keevers, M. J. Optical Properties of Intrinsic Silicon at 300 K. *Prog. Photovoltaics Res. Appl.* **1995**, *3* (3), 189–192.

(58) Dai, J.; Zhang, J.; Zhang, W.; Grischkowsky, D. Terahertz Time-Domain Spectroscopy Characterization of the Far-Infrared Absorption and Index of Refraction of High-Resistivity, Float-Zone Silicon. *J. Opt. Soc. Am. B* **2004**, *21* (7), 1379–1386.

(59) Wollack, E. J.; Cataldo, G.; Miller, K. H.; Quijada, M. A. Infrared Properties of High-Purity Silicon. *Opt. Lett.* **2020**, *45* (17), 4935–4938.

(60) Francoeur, M.; Mengüç, M. P.; Vaillon, R. Near-Field Radiative Heat Transfer Enhancement via Surface Phonon Polaritons Coupling in Thin Films. *Appl. Phys. Lett.* **2008**, *93* (4), 043109.

(61) Narayanaswamy, A.; Chen, G. Surface Modes for near Field Thermophotovoltaics. *Appl. Phys. Lett.* **2003**, *82* (20), 3544–3546.

(62) Song, B.; Fiorino, A.; Meyhofer, E.; Reddy, P. Near-Field Radiative Thermal Transport: From Theory to Experiment. *AIP Adv.* **2015**, *5* (5), 053503.

(63) Chen, D.-Z. A.; Narayanaswamy, A.; Chen, G. Surface Phonon-Polariton Mediated Thermal Conductivity Enhancement of Amorphous Thin Films. *Phys. Rev. B* **2005**, *72* (15), 155435.

(64) Biehs, S.-A. Thermal Heat Radiation, near-Field Energy Density and near-Field Radiative Heat Transfer of Coated Materials. *Eur. Phys. J. B* **2007**, *58* (4), 423–431.

(65) Miller, O. D.; Johnson, S. G.; Rodriguez, A. W. Effectiveness of Thin Films in Lieu of Hyperbolic Metamaterials in the Near Field. *Phys. Rev. Lett.* **2014**, *112* (15), 157402.

(66) Zhao, B.; Zhang, Z. M. Study of Magnetic Polaritons in Deep Gratings for Thermal Emission Control. *J. Quant. Spectrosc. Radiat. Transf.* **2014**, *135*, 81–89.

(67) Zhang, Z. *Nano/Microscale Heat Transfer*; McGraw-Hill, New York, 2007.

(68) Chapuis, P.-O.; Volz, S.; Henkel, C.; Joulain, K.; Greffet, J.-J. Effects of Spatial Dispersion in Near-Field Radiative Heat Transfer between Two Parallel Metallic Surfaces. *Phys. Rev. B* **2008**, *77* (3), 035431.

(69) Francoeur, M.; Mengüç, M. P.; Vaillon, R. Coexistence of Multiple Regimes for Near-Field Thermal Radiation between Two Layers Supporting Surface Phonon Polaritons in the Infrared. *Phys. Rev. B* **2011**, *84* (7), 75436.

(70) Mittapally, R.; Lee, B.; Zhu, L.; Reihani, A.; Lim, J. W.; Fan, D.; Forrest, S. R.; Reddy, P.; Meyhofer, E. Near-Field Thermophotovoltaics for Efficient Heat to Electricity Conversion at High Power Density. *Nat. Commun.* **2021**, *12* (1), 4364.

(71) Thurmond, C. D. The Standard Thermodynamic Functions for the Formation of Electrons and Holes in Ge, Si, GaAs , and GaP. *J. Electrochem. Soc.* **1975**, *122* (8), 1133–1141.

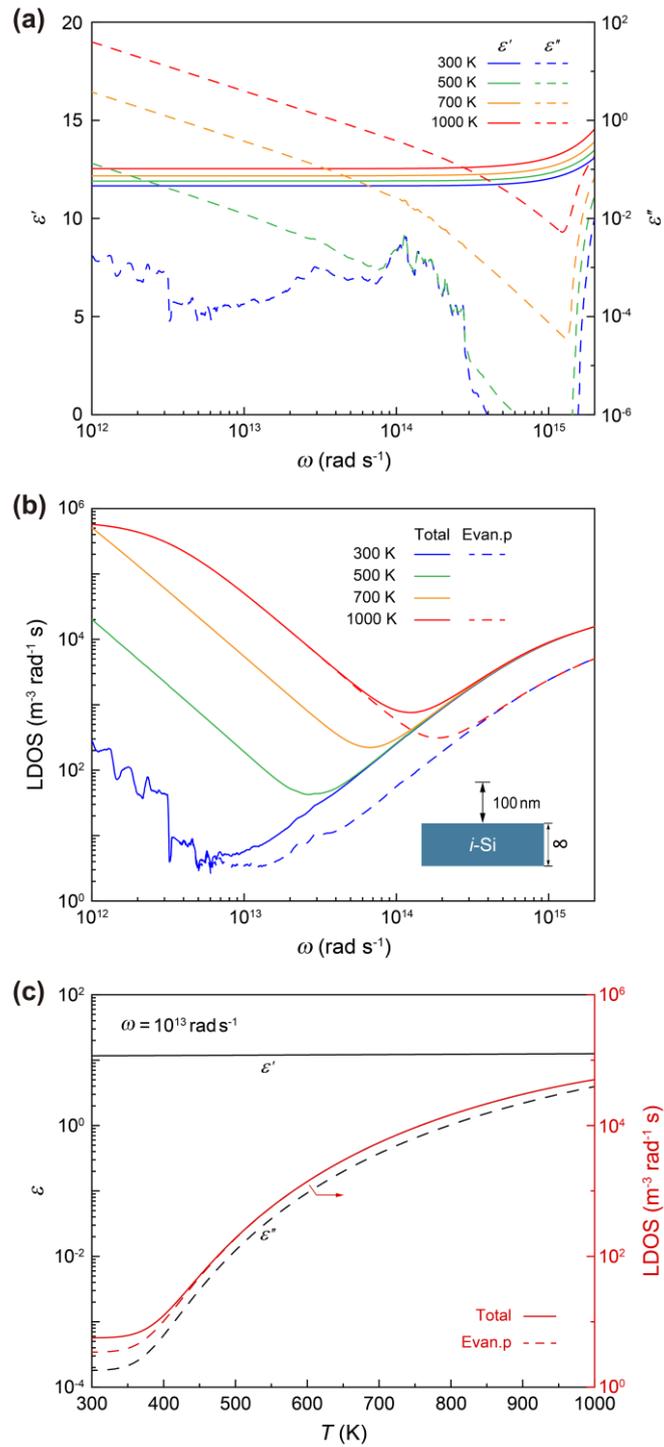

**Fig. 1.** Basic characteristics of *i*-Si. (a) Real ($\varepsilon'$) and imaginary ($\varepsilon''$) parts of the *i*-Si permittivity at different temperatures. (b) LDOS at 100 nm above bulk *i*-Si at different temperatures. (d) Permittivity and LDOS at $\omega = 10^{13}$ rad s$^{-1}$ as a function of temperature.

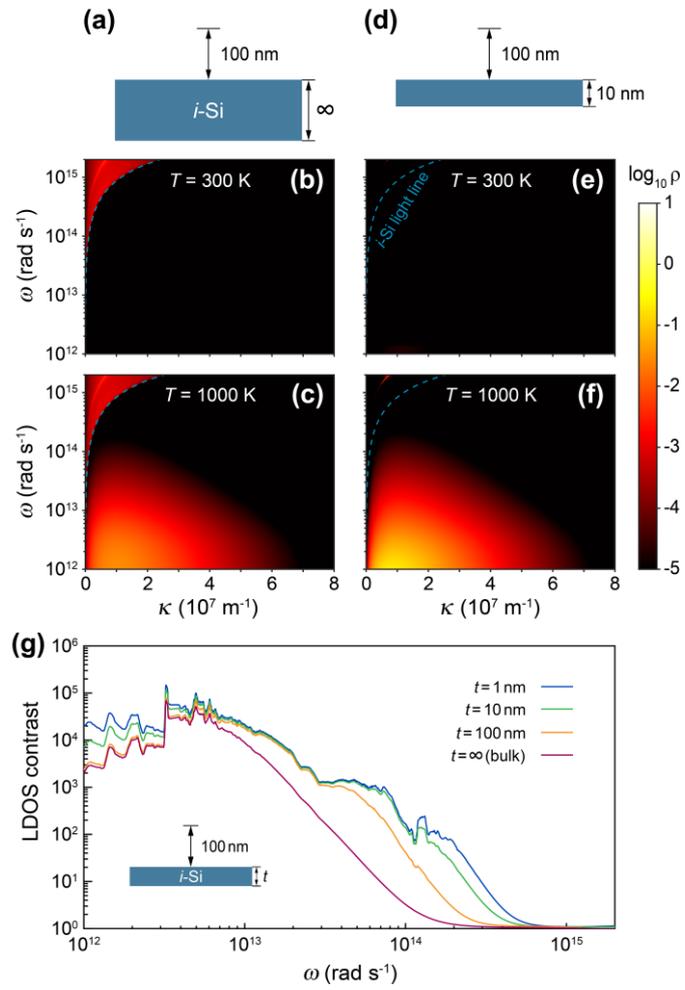

**Fig. 2.** LDOS characteristics at 100 nm above bulk and thin-film *i*-Si. (a-c) WLDOS of bulk *i*-Si, (d-f) WLDOS of thin-film *i*-Si (10 nm). (g) LDOS contrast of bulk and thin-film *i*-Si.

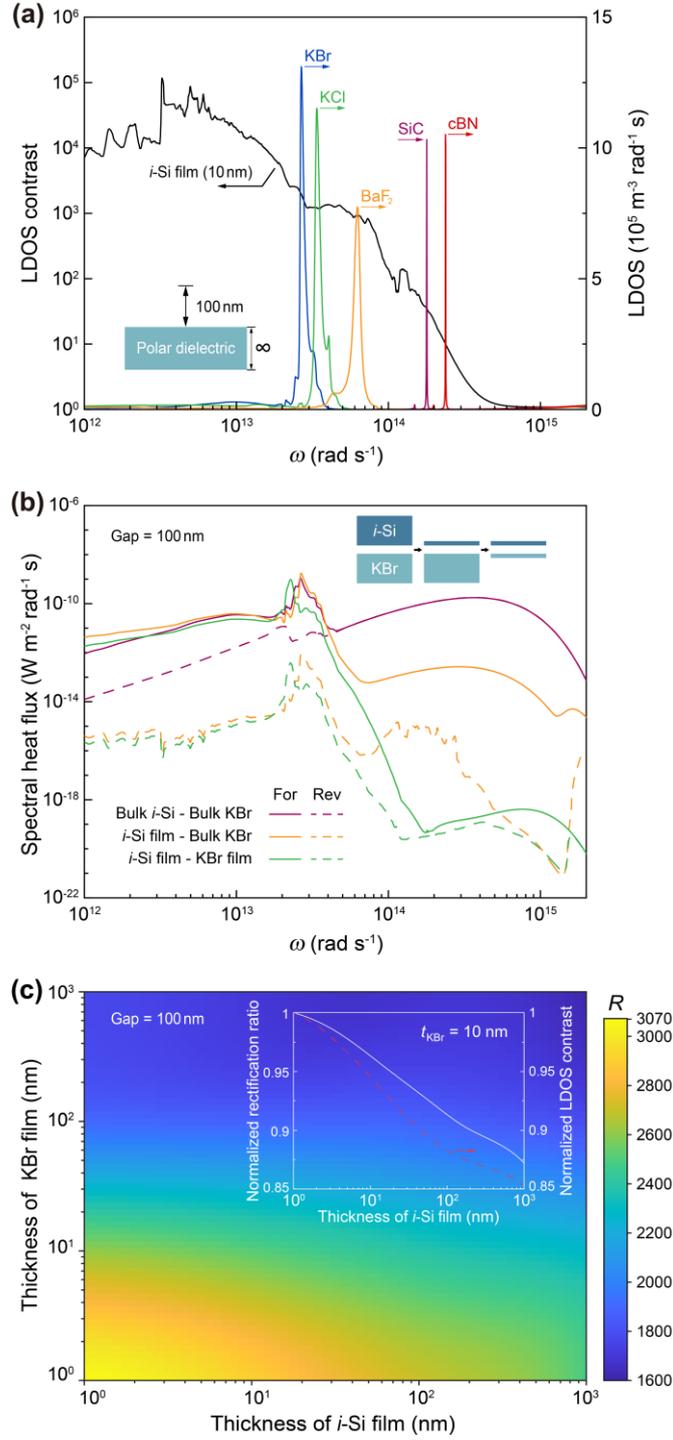

**Fig. 3.** Analysis of the diodes pairing *i*-Si with a narrowband filter (i.e., KBr). (a) LDOS contrast of a 10-nm-thick *i*-Si film and the LDOS of various bulk polar dielectrics at $z = 100$ nm. (b) Spectral heat fluxes in the forward (solid) and reverse (dashed) scenarios for different diodes at a gap of 100 nm. All the films are 10 nm thick. (c) Rectification ratio with respect to the thicknesses of the *i*-Si and KBr films at a gap of 100 nm. Inset shows the normalized rectification ratio and the normalized LDOS contrast of the *p*-polarized evanescent modes at the frequency of LDOS peak for the KBr film as a function of the *i*-Si film thickness, with the KBr film fixed at 10 nm thick.

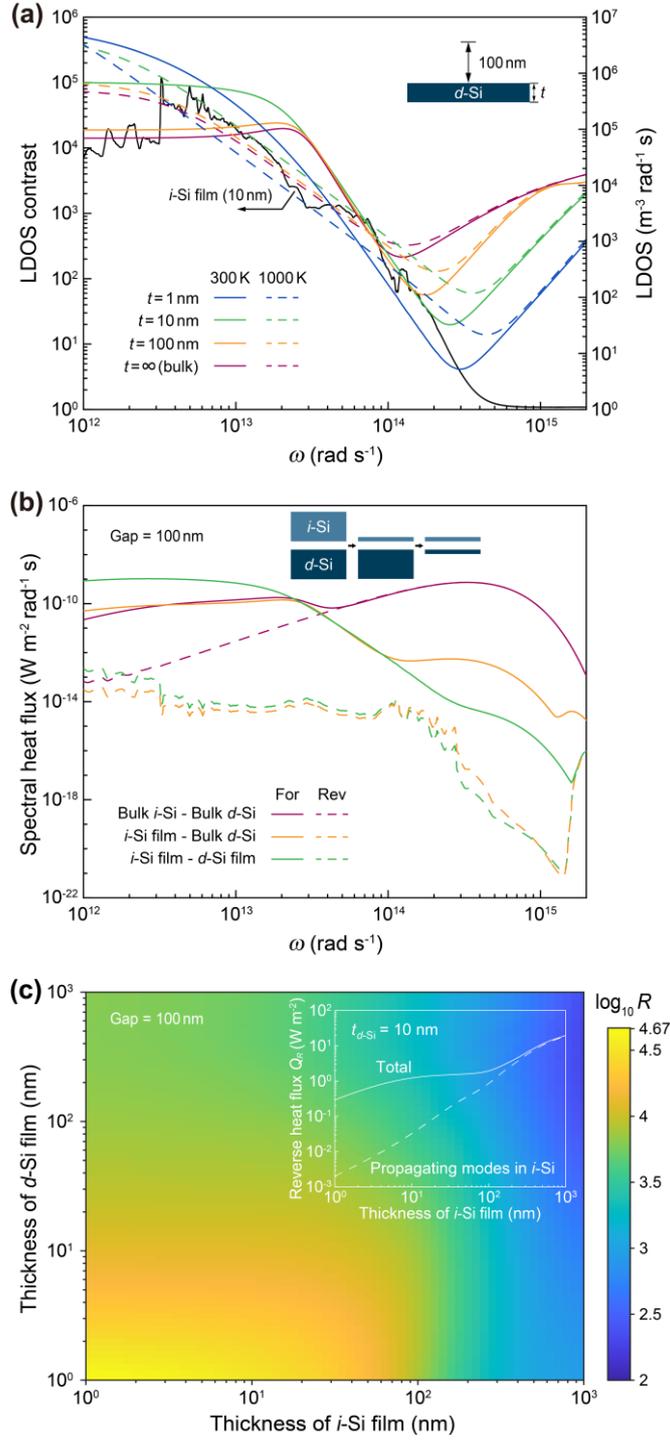

**Fig. 4.** Analysis of the diodes pairing *i*-Si with a broadband filter (i.e., *d*-Si with $N_d = 10^{18}$ cm$^{-3}$). (a) LDOS contrast of a 10-nm-thick *i*-Si film and the LDOS of bulk and thin-film *d*-Si at $z = 100$ nm. (b) Spectral heat fluxes in the forward (solid) and reverse (dashed) scenarios for different diodes at a gap of 100 nm. All the films are 10 nm thick. (c) Rectification ratio with respect to the thicknesses of the *i*-Si and *d*-Si films at a gap of 100 nm. Inset shows the reverse heat flux and the flux contributed by the propagating modes within *i*-Si as a function of the *i*-Si film thickness, with the *d*-Si film fixed at 10 nm thick.

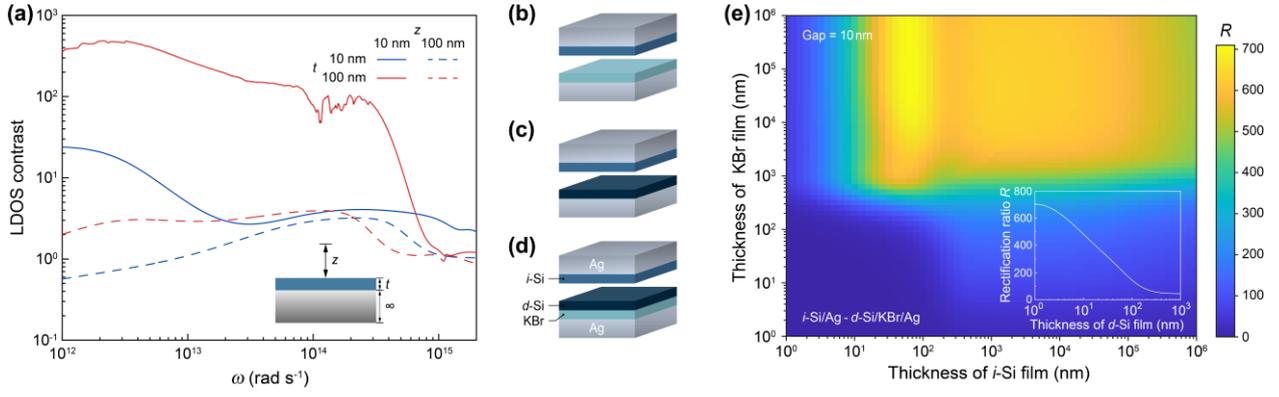

**Fig. 5.** Analysis of the multilayer diodes with Ag substrates. (a) LDOS contrast for an *i*-Si film sitting on a bulk Ag substrate. (b) Schematic of the diode pairing *i*-Si/Ag with KBr/Ag. (c) The pair of *i*-Si/Ag and *d*-Si/Ag. (d) The pair of *i*-Si/Ag and *d*-Si/KBr/Ag. (e) Rectification ratio for the pair of *i*-Si/Ag and *d*-Si/KBr/Ag with respect to the thicknesses of the *i*-Si and KBr films, at a gap of 10 nm. The doping level and thickness of *d*-Si film are the optimized values of $6\times10^{18}$ cm$^{-3}$ and 1 nm (Table S1). Inset shows the rectification ratio as a function of the *d*-Si film thickness, with the thicknesses of the *i*-Si and KBr films fixed at the optimal values of 67 nm and 129 µm, respectively (Table S1).

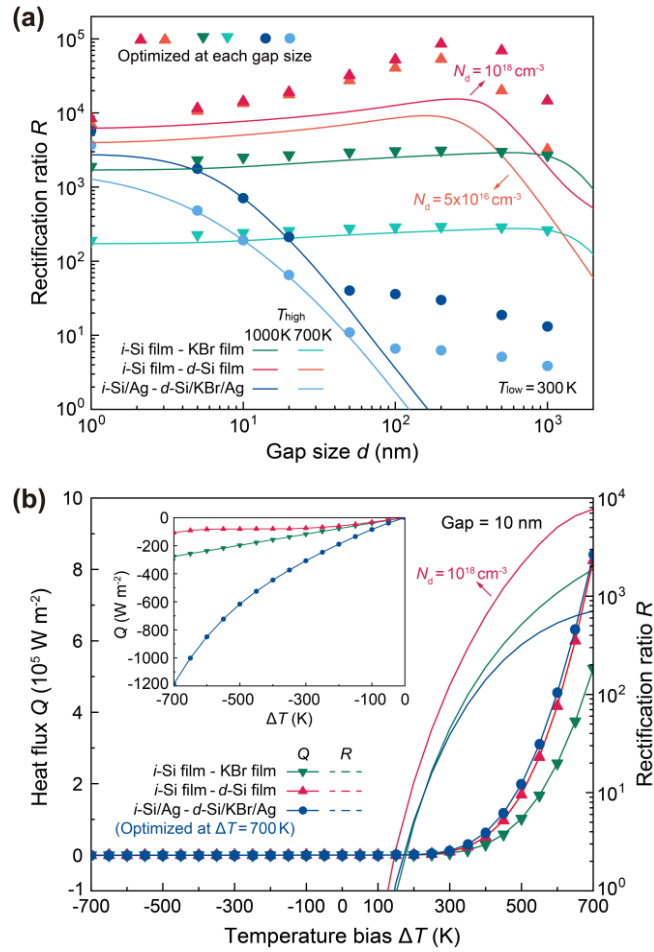

**Fig. 6.** Rectification performance of representative *i*-Si-based diodes. (a) Gap-dependent rectification ratios. (b) Heat fluxes and the corresponding rectification ratios under different temperature biases at a gap of 10 nm. The temperature bias denotes $\Delta T = T_1 - T_2$ where $T_1$ is the terminal temperature of the *i*-Si side and $T_2$ that of the opposite side. The lower temperature is fixed at 300 K. All the film thicknesses for the suspended-film diodes are 10 nm. For the multilayer diode, the parameters are optimized at a 10-nm gap (Table S1).